\documentclass[epj,amsfonts]{svjour}
\usepackage{graphics}
\usepackage{graphicx}% Include figure files
\usepackage{dcolumn}% Align table columns on decimal point

\newcommand{\ar}{\arrowvert}

\newcommand{\cd}{\! \cdot \!}
\newcommand{\be}{\begin{equation}}
\newcommand{\ee}{\end{equation}}
\newcommand{\ba}{\begin{eqnarray}}
\newcommand{\ea}{\end{eqnarray}}

\begin{document}
\title{From Euclidean to Minkowski space \\ 
with the Cauchy-Riemann equations}
\author{ 
Mercedes Gimeno-Segovia and Felipe J. Llanes-Estrada
}                     % Do not remove
\institute{ Departamento de F\'{\i}sica Te\'orica I,  Universidad
Complutense, 28040 Madrid, Spain }
\date{Received: date / Revised version: date}

\abstract{ We present an elementary method to obtain Green's 
functions in non-perturbative quantum field theory in Minkowski space 
from calculated Green's functions in Euclidean space. Since in
non-perturbative field theory the analytical structure of amplitudes
is many times unknown, especially in the presence of confined fields,
dispersive representations suffer from systematic uncertainties.
Therefore we suggest to use the Cauchy-Riemann equations, that
perform the analytical continuation without assuming global
information on the function in the entire complex plane, only
in the region through which the equations are solved.
We use as example the quark propagator in Landau gauge Quantum 
Chromodynamics, that is known from lattice and Dyson-Schwinger studies 
in Euclidean space. The drawback of the method is the instability of the 
Cauchy-Riemann equations to high-frequency noise, that makes
difficult to achieve good accuracy.
We also point out a few curiosities related to the Wick rotation.
 % end of PACS codes
} %end of abstract
\authorrunning{Gimeno-Segovia and Llanes-Estrada}
\titlerunning{Wick-rotation and Cauchy-Riemann}
\maketitle

%%%%%%%%%%%%%%%%%%%%%%%%%%%%%%%%%%%%%%%%%%%%%
\section{The Wick rotation}
%%%%%%%%%%%%%%%%%%%%%%%%%%%%%%%%%%%%%%%%%%%%%

\begin{figure}[htbp]
\includegraphics[width=7cm,angle=-90]{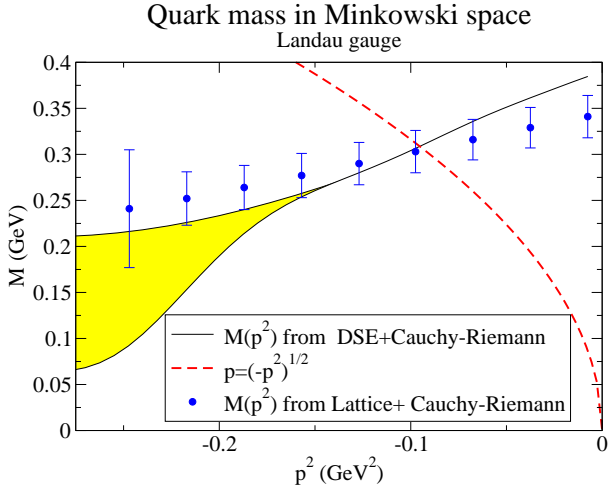}
\caption{The analytical continuation to negative Euclidean squared
momentum (positive quark virtuality in Minkowski space) of the Landau
gauge quark mass function. The yellow band is obtained with input from
an Euclidean Dyson-Schwinger calculation, where the Cauchy-Riemann
equations have been solved with the $\theta$ method in different grids.
The Lattice data is likewise analytically continued with the
Cauchy-Riemann equations. At low $p^2$ the error is dominated by the
original lattice statistical error, eventually the build-up of numerical
errors in the Cauchy-Riemann equations dominate.
The trend of $Re(M)$ is clearly decreasing with $p^2$, and therefore a
pole in the quark propagator is expected. Since $Im(M)$, presented later
on, is small, this pole is on or close to the real axis, with 
$M=305(25)\ MeV$. The lattice data that is analytically continued is 
from ref. \cite{Parappilly:2005ei}.
\label{fig:minkowski_mass}}
\end{figure}

Central to non-perturbative quantum field theory is the computation of
Green's functions, the vacuum expectation value of quantum operators.
These and the related Scattering Matrix elements are most often computed
in Euclidean space, defined by the transformation
\be
t\to -it_E \ \ \ \   k_0 \to ik_0 \  .
\ee
This coordinate transformation is known as ``Wick rotation'' (see for 
example \cite{O'Brien:1974mr} for a short account). There are many advantages
in solving the field equations in terms of the rotated variables to obtain
so called Schwinger functions, and
we list some below in section \ref{sec:whywick}. 

Once the wanted functions have been computed in Euclidean (momentum) space
$k_E=(ik_0,{\bf k})$ one would wish to recover the original Minkowski
space Green's functions by inverting the Wick rotation. This is possible
in perturbation theory at low orders \cite{peskin,itzykson}  where one
has explicit expressions for the functions, and their analytical structure
(poles, cuts, essential singularities) is at hand, so that one can employ
Cauchy's theorem and collect dispersive cut integrals or
pole residues if need be, and obtain the Minkowski space Green's function
by analytical continuation. 

For example, to obtain the electron propagator in momentum space
(Fourier transform of the probability amplitude for the
electron to reach point $x$ if it was originally at the origin $0$) 
\be
S(p) = \int d^4 x e^{-ix\cd p} S(x,0)
=i Z(p^2) \frac{\not p + M(p^2)}{p^2-M^2(p^2)+i\epsilon}
\ee
one would perform the Wick rotation $p_0\to ip_{0E}$
and obtain a function $S(p_E)$ as a perturbation of its free-field values
($M(p^2)=m$, $Z(p^2)=1$). With the function explicitly known, one
just extends it into the complex plane and simply substitute its argument
$p_{0E}$ by $-ip_0$. If the function is not totally known but its analytical
structure is, one employs Cauchy's theorem as mentioned.

However, in non-perturbative quantum field theory one is seldom in this
desirable situation. More often than not, the function has been calculated
with the help of a computer, 
be it by solving the Dyson-Schwinger equation
\cite{dseroberts} with some carefully designed truncation, or by 
trying
a Montecarlo evaluation averaging over a small number of configurations
on a lattice \cite{Creutz:2000bs}. 

The outcome is that the function is then known for Euclidean momenta,
typically $k_E^2>0$, and an extension into the complex plane becomes
necessary to reach the negative axis $k_E^2<0$, $k_E^2=-k^2$. 
For non-perturbative
functions one sometimes ignores the precise analytic structure in the complex
plane. This situation is worsened in theories where the field quanta 
do not appear in asymptotic states, except in very specific 
combinations, such as is presumably the case for Quantum 
Chromodynamics.

Attempts have of course been made to solve the problem. An obvious 
approach is to perform
a theory-motivated fit to the computer data, building-in well educated
guesses on what the analytical structure of the continued function
must be \cite{Alkofer:2003jj,Alkofer:2004cw}.

Another possibility is to write a spectral representation in terms of 
a Stieltjes transform yielding a spectral density $\rho$, 
\be
S(p^2) = \int dp \frac{\rho(m^2)}{p^2-m^2+i\epsilon}
\ee
(we have eliminated spin structure)
and then solving the Dyson-Schwinger equation directly ``in Minkowski
space'' for $\rho$, finally inverting the Stieltjes transform to recover
the propagator or Bethe-Salpeter wanted function \cite{Sauli:2006ba},
\cite{Kusaka:1995za}.
It is of course
clear that in writing the spectral representation, one is already assuming
a given cut structure for the function.
It is known that, for local quantum field theories where the quanta 
appear in the final state, the propagator in coordinate space in 
the upper-half $x$
complex plane is analytic \cite{itzykson}, but for 
confined quanta many questions remain.

In this paper we enrich the toolbox by putting forward a very simple 
method that does not require the function's analytical structure to be 
known on the entire complex plane. We observe that the analyticity of a 
given function cannot only be formulated globally, through satisfaction 
of Cauchy's theorem, but also locally, through satisfaction of the 
Cauchy-Riemann equations. By integrating this simple first order 
differential system with initial condition the function in a given 
region computed by other means (DSE, lattice, exact renormalization group 
equations \cite{Pawlowski:2005xe}, \cite{Parappilly:2005ei}, etc.) one 
can 
achieve two goals. First,
the numerically obtained solution can break down at a given point or 
line in the plane, indicating perhaps a pole or other singularity. 
Second, if the region where the system is integrated avoids such 
singularities, one can obtain the analytical extension (within errors 
and non-uniqueness) to another region in the complex plane.

We use as example the behavior of the quark mass function in 
Landau gauge QCD, $M(p^2)$, that we analytically continue from 
positive Euclidean virtuality ($p^2>0$) into Minkowski space 
with ($p^2<0$).  This is plotted in figure \ref{fig:minkowski_mass}
It can be seen that, within the statistical errors inherited 
from lattice data and the systematic numerical errors intrinsic to our 
procedure, the mass function decreases with increasing Minkowski $p^2$. 
Our result keeps open the possibility of a pole of the quark propagator 
for real $p^2$ (whose absence has been at times thought as a possible 
sign of confinement).

The rest of the paper consists of four sections. In section \ref{sec:whywick}
we make a few comments, some common place but others quoted less often,
about the advantages of initially working in Euclidean space. In section
\ref{sec:numeric} we present the Cauchy-Riemann method with one practical
case, the quark propagator.
A few theoretical comments about errors involved in the process and the
generalization to more dimensions are left for section 
\ref{sec:theory}. Our discussion is summarized in section \ref{sec:conclusions}.

%%%%%%%%%%%%%%%%%%%%%%%%%%%%%%%%%%%%%%%%%%%%%%%%%%%%%%%%%%%%%%%%%%%%%%
\section{Working in Euclidean space \label{sec:whywick}}
%%%%%%%%%%%%%%%%%%%%%%%%%%%%%%%%%%%%%%%%%%%%%%%%%%%%%%%%%%%%%%%%%%%%%%
In lattice formulations of Quantum Field Theory, the field 
configurations over which the path integral is evaluated are randomly
generated according to a distribution 
$e^{-\int dt {\mathcal L}}$, that is the Wick-rotation of the actual 
quantum weight for the path integral, $e^{i\int dt {\mathcal L}}$. 
Thereafter computed lattice  Green's functions are valid in Euclidean 
space.
Even if working in Minkowski space, a popular way
of ``minimally'' regularizing in the path integral formalism is to
rotate the time integration into the complex plane
$$
\int d^4 x = \lim_{T\to\infty} \int_{-T}^T dt \int d^3 x \ .
$$

Beyond the convergence of the path integral and the weighting 
configurations inside a compact set of function space, there are 
several more advantages. 

One is that the Dyson-Schwinger equations, whose solutions are often 
used to interpret lattice data, are extremely difficult to solve on a 
computer in Minkowski space. 
Indeed, a typical DSE is that for the mass function of a fermion in the 
presence of a scalar field, with Yukawa coupling, in rainbow 
approximation
\be
M(p)= c \int d^4k \frac{M(k)}{(k^2-M(k)^2+i\epsilon)
((k-p)^2-m_\phi^2+i\epsilon)} \ .
\ee
If one solves the equation iteratively by guessing $M_0$, one notices 
that the position of the fermion pole in the denominator is not known 
after the first iteration, and it needs to be determined numerically (an 
attempt at carrying on this program exists \cite{Bicudo:2003fd}).

However the common use is to Wick-rotate the $k$ integration variable to 
Euclidean space. If $p$ is likewise rotated, the resulting equation is 
easier to program as the denominator poles are on the left $k^2$ plane, 
out of the numerical integration region in the radial $k^2$ variable.
\be
M(p)= c \int d^4k \frac{M(k)}{(k^2+M(k)^2+i\epsilon)
((k-p)^2+m_\phi^2+i\epsilon)} 
\ee
($c$ represents constants irrelevant to the discussion).

Note also that when working in Euclidean space, a discrete subgroup of 
the Euclidean rotation group is retained. For example, for a simple 
cubic lattice with $(x=ai_x,y=ai_y,z=ai_z,t=ai_t)$, invariance under 
rotations by $\pi/2$ is explicit. This can be exploited to study the 
quantum representations of the discrete group, then trying to match the 
resulting states to a representation of the full continuous group.

However, in Minkowski space there is no finite lattice that retains 
invariance under a non-trivial subgroup of the Lorentz group. If a grid 
is invariant under discrete Lorentz transformations of parameter $a$, 
then it has infinitely many points. We discuss Lorentz-invariant 
discretizations of Minkowski space in the appendix.

One further motivation is the non-compactness of the equal $k^2$ 
hypersurfaces. While in Euclidean space the condition 
$k^2=\Lambda^2$ determines a hypersphere's surface, so that
\be
\int^{\Lambda} d^4k_E f(k_E^2) = 2\pi^2 \int^\Lambda k_E^3dk_E 
f(k_E^2)
\ee
can be factorized into a radial integral and a finite $2\pi^2$ solid 
hyperangle (the hyperarea of a unit-radius hypersphere's surface)
 this is not possible in 
Minkowski space, where the corresponding unit-hyperboloid
$k^2=1$ has infinite hypersurface.
Therefore, integrals of Lorentz-invariant functions are by necessity 
divergent even after regulation of large virtualities, and are only 
defined by analytical continuation from Euclidean space. 

In perturbation theory, a much used method is to perform the $k_0$ 
integrals first, usually with pole analysis, and later impose a cutoff 
on spacelike momentum $\mathbf{k}$. However this cutoff is 
frame-dependent, and there is no direct method that manifestly preserves 
Lorentz invariance.

From all these arguments, it is hard to conceive
progress in non-perturbative quantum field-theory in Minkowski space 
without progress in complex-plane analytical continuation for the 
relevant functions. This paper is a modest contribution in this 
direction, with the interest in keeping the discussion alive.

%%%%%%%%%%%%%%%%%%%%%%%%%%%%%%%%%%%%%%%%%%%%%
\section{Numerical solution of the Cauchy-Riemann equations
\label{sec:numeric}}
%%%%%%%%%%%%%%%%%%%%%%%%%%%%%%%%%%%%%%%%%%%%

%%%%%%%%%%%%%%%%%%%%%%%%%%%%%%%%%%%%%%%%%%%%
\subsection{Cauchy-Riemann equations in polar coordinates}
%%%%%%%%%%%%%%%%%%%%%%%%%%%%%%%%%%%%%%%%%%%%

If $u$ and $v$ are respectively the real and imaginary parts of a 
complex 
function of one complex variable $p^2=re^{i\theta}$, the Cauchy-Riemann
equations in polar coordinates read
\ba
\frac{\partial v}{\partial \theta} = r \frac{\partial u}{\partial r}
\\ 
\frac{\partial u}{\partial \theta}= -r \frac{\partial v}{\partial 
\theta} \ .
\ea 

\begin{figure}[htbp]
\includegraphics[width=6cm]{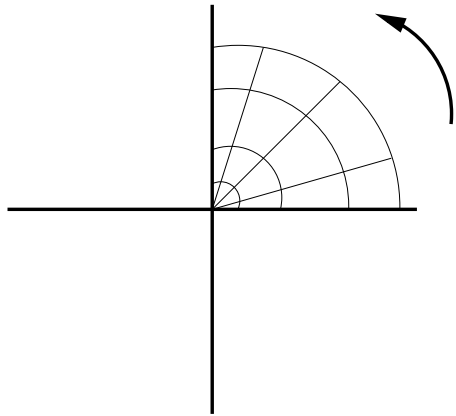}
\caption{The Cauchy-Riemann equations in polar coordinates allow to 
explore a fan-shaped region of the complex plane where a function is
analytic.
 \label{fig:polar}}
\end{figure}

Given the initial conditions $u(r,0)=u_0(r)$, $v(r,0)=v_0(r)$
 on a segment of the real $p^2$ axis, corresponding to
$\theta=0$, one can then evolve the system towards increasing and 
decreasing
$\theta$ (like the opening of a fan). For very smooth data sets one can
typically reach 90-120 degrees
on each side of the fan before the instabilities wipe the solution to 
infinity.
The (Cauchy-Euler) explicit discretization  with centered 
$r$-derivative on a grid $(r_j,\theta_i)$ is simply
\ba \nonumber
v(r_j,\theta_{i+1})= v(r_j,\theta_i)+r_j (\theta_{i+1}-\theta_i)
\frac{u(r_{j+1},\theta_i)-u(r_{j-1},\theta_i)}{r_{j+1}-r_{j-1}}
\\  \\ \nonumber
u(r_j,\theta_{i+1})= u(r_j,\theta_i)-r_j (\theta_{i+1}-\theta_i)
\frac{v(r_{j+1},\theta_i)-v(r_{j-1},\theta_i)}{r_{j+1}-r_{j-1}}
\\
\ea
where, to solve over an arch taken anticlockwise, 
$(\theta_{i+1}-\theta_i)>0$. 
At the end-points of the grid one cannot use centered derivative, so 
left (right derivative) is necessary, 
$(v(r_2,\theta_i)-v(r_1,\theta_i))/(r_2-r_1)$, etc.
The situation is represented in figure
\ref{fig:polar}.

The use of this method is to provide a cross-check of solutions of the 
Dyson-
Schwinger equations in the complex plane. These are needed to solve
the Bethe-Salpeter equations for mesons, since the (external) 
meson momentum is of course in Minkowski space (real), and the 
internal quark momentum is Wick-rotated to Euclidean space (imaginary)
so that one ends solving the DSE inside a parabolla in the complex plane
symmetric respect to the real momentum axis. The Cauchy-Riemann 
equations
are currently no match in precision to directly solving the DSE in 
the complex plane where this is feasible, but they can provide a cross
check that is very simple to programme (compare the trivial linear 
system above with the complex, non-linear, bidimensional DSE when the
angular kernel or vertex are non-trivial).

\begin{figure}[h]
\includegraphics[width=7cm,angle=-90]{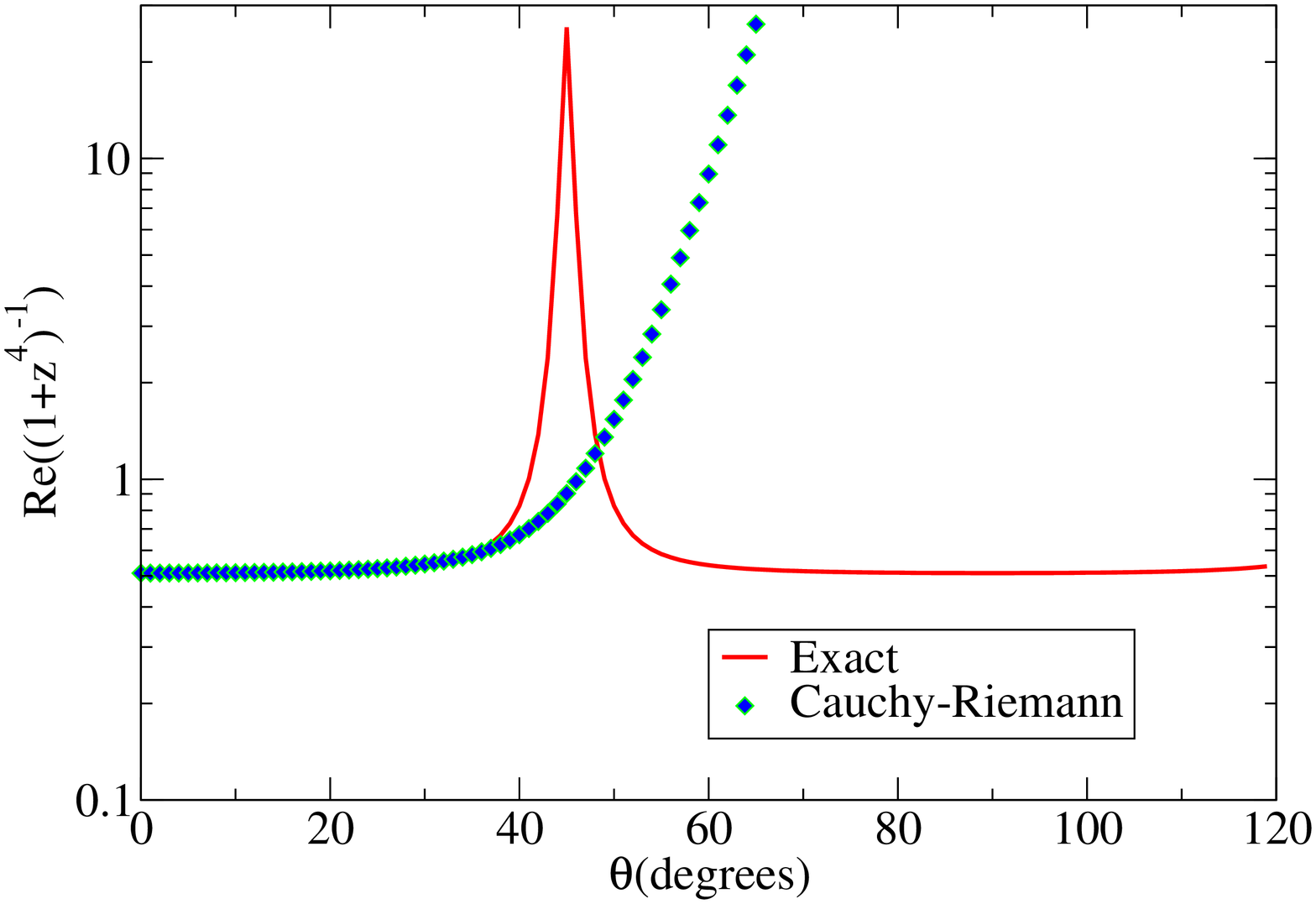}
\includegraphics[width=7cm,angle=-90]{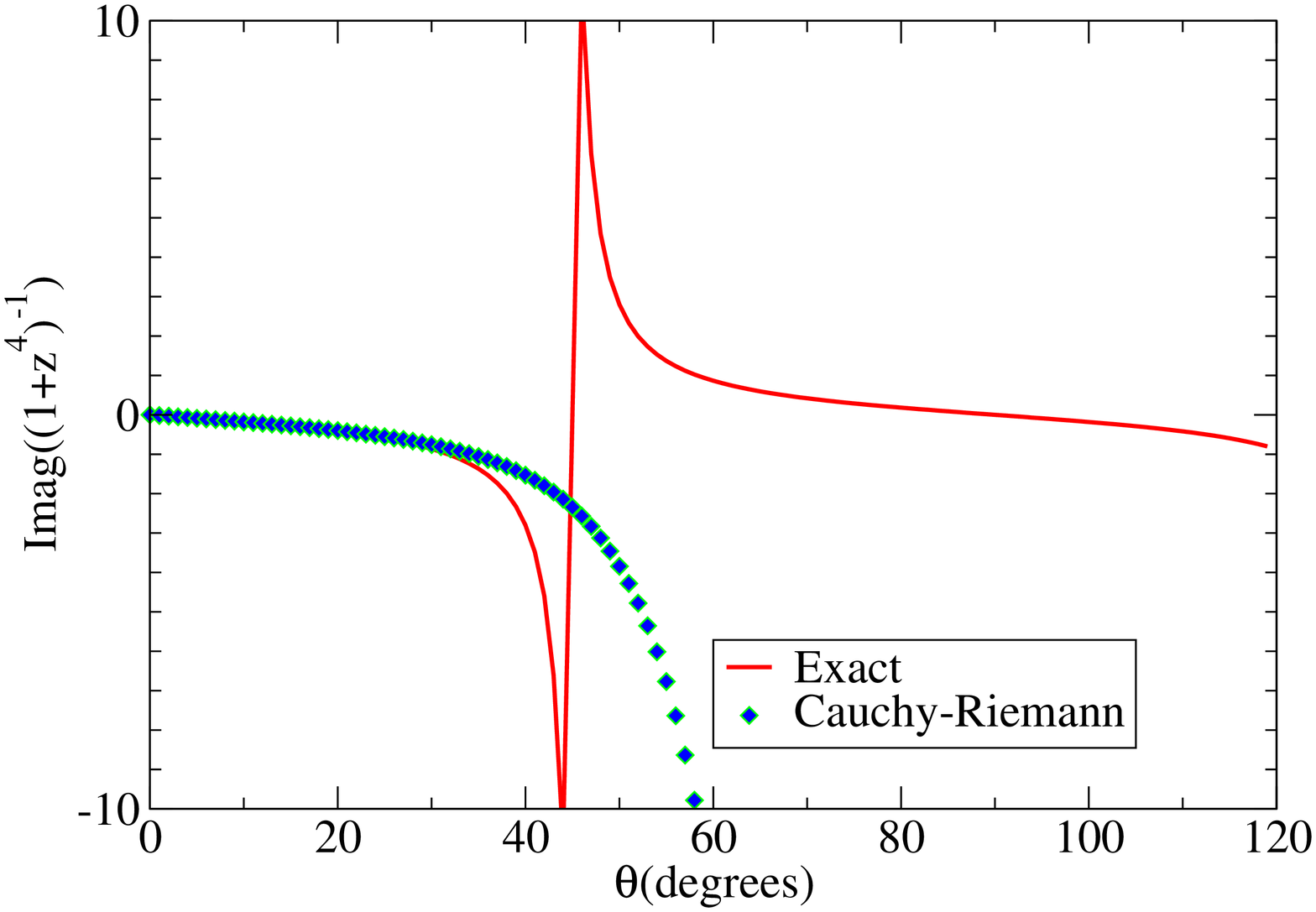}
\caption{ Real and imaginary part of the function $\frac{1}{1+z^4}$ for 
$z=re^{i\theta}$ with fixed $r=0.99$ and varying $\theta$. The dashed 
line is the computer solution of the Cauchy-Riemann equations in degree 
steps, with the initial condition given on the positive real half-axis. 
The function has a pole at $(1+i)/\sqrt{2}$ that the Cauchy-Riemann 
equations cannot isolate since they entail analyticity. They however 
diverge due to accruing instability in the region where the function 
has larger derivative.
\label{fig:poleCR}
}
\end{figure}

The Cauchy-Riemann equations however are a statement of analyticity,
and the solution is a numerical representation of the closest analytical
function that contains the initial data.  This means that if the 
``true'' function has a pole or a cut, the Cauchy-Riemann iteration
will fail to see it and simply separate from that function, and is 
likely
to diverge soon from accruing instabilities. This is illustrated in 
figure
\ref{fig:poleCR}. 

%%%%%%%%%%%%%%%%%%%%%%%%%%%%%%%%%%%%%%%%%%%%%%%%%%%%%%%%%%%%%
\subsection{Cauchy-Riemann equations on a strip}
%%%%%%%%%%%%%%%%%%%%%%%%%%%%%%%%%%%%%%%%%%%%%%%%%%%%%%%%%%%%%

\begin{figure}[htbp]
\includegraphics[width=6cm]{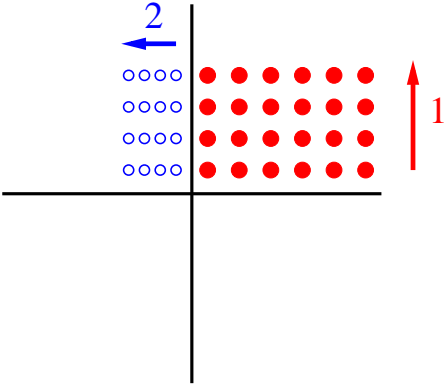}
\caption{The Cauchy-Riemann equations in Cartesian coordinates can be 
used to explore a strip, first upwards along the $y$axis, then leftwards 
along the $x$ axis. In this use one only needs analyticity of the 
function on a strip above (or below) the axis to obtain information on 
the Minkowski side. The method fails only if the complex plane is 
completely cut from $-\infty$ to $\infty$ along the $y$ axis. 
 \label{fig:cartesian}}
\end{figure}

The advantage of a local formulation of analyticity employing the 
Cauchy-Riemann equations is lost if one needs to swipe the entire 
complex plane. Therefore it is profitable to solve them in 
Cartesian coordinates, first away from the $x$ axis along $y$
\ba
\frac{\partial u}{\partial y} =- \frac{\partial v}{\partial x}\\
\frac{\partial v}{\partial y} =  \frac{\partial u}{\partial x} 
\ea
and then leftwards along $x$. 
The method fails if the complex plane is completely
cut from $-\infty$ to $\infty$ along the $y$ axis. In any other 
situation
(the standard half-plane cut of a power-law or logarithm, or a finite
number of poles or essential singularities), one can find a path between
the right and the left $x$ axis and solve the Cauchy-Riemann equations
along them.

We now improve upon the discretization of the differential equations and 
employ an implicit $\theta$ method. This is convenient since the 
Cauchy-Riemann are quite unstable (as pointed out below in subsection 
\ref{subsec:stab}).
The $\partial_y u$ equation for a point not on the edge of 
the grid becomes (with the superindex labelling $y$, the subindex  $x$)
\ba
\frac{u_i^{j+1}-u_i^j}{y_{j+1}-y_j} = 
\frac{-1}{x_{i+1}-x_{i-1}} \times \\ \nonumber
\left(
\theta (v_{i+1}^{j+1}-v_{i-1}^{j+1}) +(1-\theta) (v_{i+1}^{j}-v_{i-1}^{j})
\right)\ .
\ea 

In the advance along $y$ one groups the $u_i$ and $v_i$ for fixed $y_j$ 
in a vector ${\mathbf{u}}^{j}=(u_1^j,v_1^j,u_2^j,v_2^j,...,u_N^j,v_N^j)$ 
and the $\theta$ method's 
discretization can be written down as a linear problem
\be
{\mathbf{A}} {\mathbf{u}}^{j+1} = {\mathbf{B}} {\mathbf{u}}^{j}  \ .
\ee

In the simplest case of equal $x$ subintervals, one can define 
$r=\theta \Delta y/(2\Delta x)$, then the matrix $\mathbf{A}$ becomes
\be
\!\! \left(\! \! \! 
\begin{tabular}{cccccccccc}
$1$  & $-2r$ & $0$   & $2r$ & $0$ & $\dots$ &        & & & \\
$2r$ & $1$   & $-2r$ & $0$  & $\dots$ &         &        & & & \\
$0$  & $-r$  & $1$   & $0$  & $0$ & $r$       &$\dots$ & & & \\
$r$  & $0$   &  $0$  & $1$  &$-r$    & $0$     &$\dots$ & & & \\
$\cdots$ &   &       &      &         &         &        & & & \\
  & & & & & $\dots$ & $0$ &$-2r$ & $1$& $2r$\\
               & & & & $\dots $ & $0$ & $2r$ & $0$ & $-2r$  & $1$\\
\end{tabular}\!\!\!
\right)\!
\ee
where the third and fourth line are the repeated unit, except the 
non-vanishing elements are shifted to the right (the $1$'s always mark 
the diagonal, and the matrix is band-diagonal).
The matrix $\mathbf{B}$ can likewise be filled by exchanging $\theta\to 
 (1-\theta)$ and changing the sign of all off-diagonal matrix elements 
of $\mathbf{A}$ (the diagonal of $\mathbf{B}$ likewise contains $1$'s).

The advance in $y$ proceeds by solving the linear system to obtain 
$(u,v)$ at $y_{j+1}$ from their values at $y_j$. We use standard $LU$ 
factorization, although since the matrix $\mathbf{A}$ is band-diagonal, 
Crout's algorithm can speed things some. For one complex variable and 
the small number of points we use this is irrelevant in computer time.

However, since the advance to the left will have as initial condition 
the edge (along the $y$ axis) of the first computed block, and at the 
end we will be interested on the values of the function on the $x$ axis, 
that is the edge of the second computed block, it pays off to improve 
the computation of the derivative at the edge of the block.

The left-derivative we have displayed explicitly,

\be
\frac{\partial u(x=0,y)}{\partial x} = 
\frac{u_2^j-u_1^j}{x_2-x_1}+o(h)
\ee
can be interpreted as a centered derivative at the mid-point 
$(x_2+x_1)/2$. Considering also the centered derivative at $x_2$ and 
extrapolating linearly to $x_1$, one obtains an improved
\be
\frac{\partial u(x=0,y)}{\partial x} = 
2 \frac{u_2^j-u_1^j}{x_2-x_1} - \frac{u_3^j-u_1^j}{x_3-x_1} +o(h^2) \ ,
\ee
and likewise at the last $N$ point of the grid, and this slightly 
complicates the first two and last two rows of $\mathbf{A}$.

Once the advance upwards in the $y$ direction has reached $N$, one 
starts an advance to the left as in figure \ref{fig:cartesian} and 
similar considerations apply.

The $\theta$ parameter that advances or delays the derivative 
perpendicular to the integration direction is empirically fixed for now. 
Several problems are somewhat independent of $\theta$, others have a 
broad minimum of instability around -1.  We find that 
$\theta=-0.5$ is as good as any. A brief eigenvalue analysis is 
presented below that explains why, in subsection \ref{subsec:stab}.

\begin{figure}[htbp]
\includegraphics[width=7cm,angle=-90]{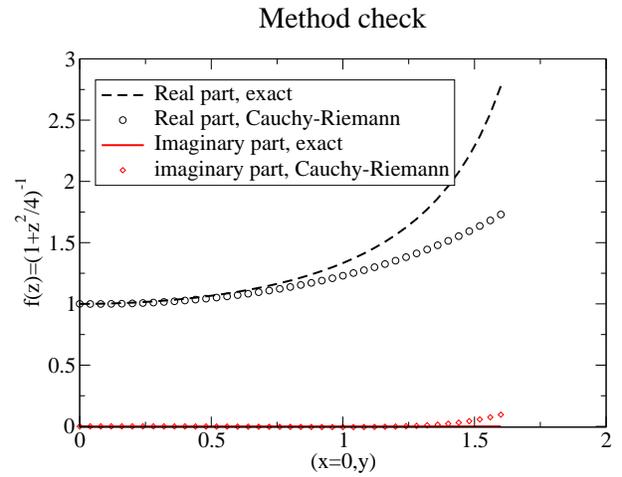}
\caption{\label{fig:rationaltesty}
The function $f_{\rm test}(z)= \frac{1}{1+z^2/4}$ analytically continued 
from $(x>0, y=0)$ to $(x=0,y>0)$ with the Cauchy-Riemann equations. 
The imaginary part is exactly zero on 
this imaginary axis, the function has a pole at $y=2$. 
}
\end{figure}

\begin{figure}[htbp]
\includegraphics[width=7cm,angle=-90]{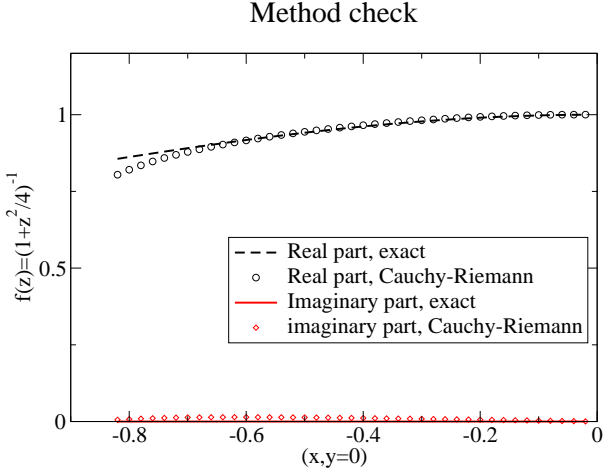}
\caption{\label{rationaltestx}
The function $f_{\rm test}(z)= \frac{1}{1+z^2/4}$ analytically continued
from $(x>0, y=0)$ to $(x<0,y=0)$ with the Cauchy-Riemann equations.
}
\end{figure}

To show a test of the method, we employ the simple function
\be
f_{\rm test}(z)= \frac{1}{1+z^2/4}
\ee
that has two poles above and below the $x$ axis at $\pm 2i$. Because of 
the fast build-up of numerical errors, we need $\Delta y<\Delta x$ so 
the strip is always shorter in the direction of the advance of the 
integration (for an $N\times N$ problem). 
In figs. \ref{fig:rationaltesty} and \ref{rationaltestx} we show 
the real and imaginary parts calculated with the Cauchy-Riemann 
equations with initial condition on the positive real half-axis, plotted 
along the imaginary and the negative real-axis respectively. 
\begin{figure}[htbp]
\includegraphics[width=7cm,angle=-90]{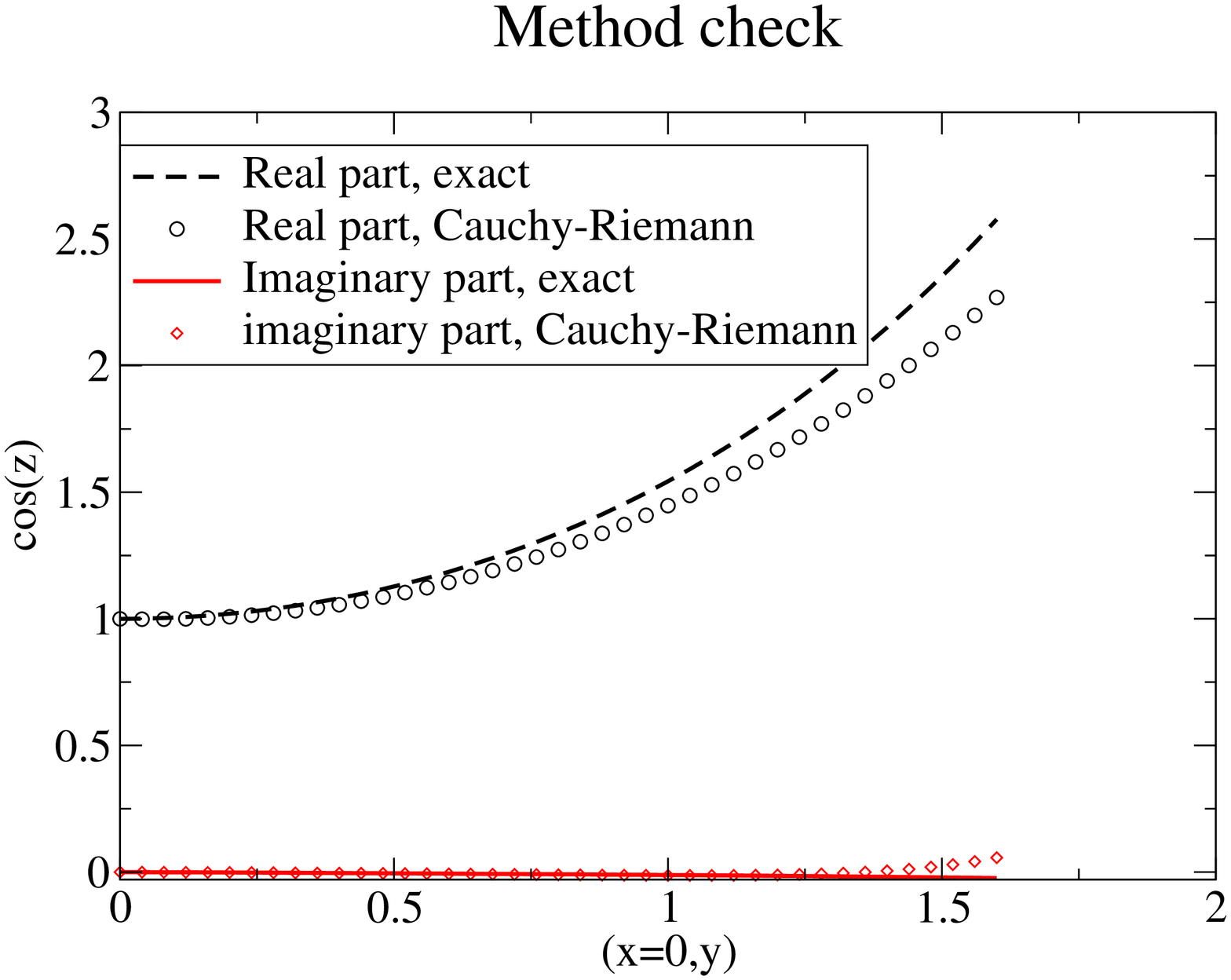}
\caption{\label{costesty}
As in figure \ref{fig:rationaltesty} but for the function $\cos z$.
}
\end{figure}
\begin{figure}[htbp]
\includegraphics[width=7cm,angle=-90]{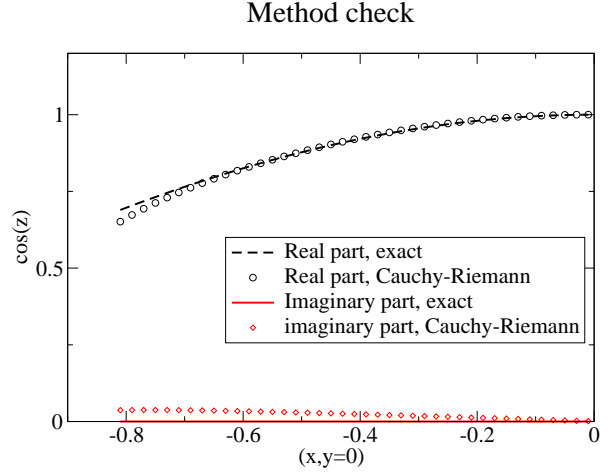}
\caption{\label{costestx}
As in figure \ref{rationaltestx} but for the function $\cos z$.
}
\end{figure}

The method starts breaking 
down for $x$ near zero and $y>1$ as the pole is approached. But one can 
see how the part of the strip that goes well below the pole passes 
cleanly and allows to reproduce the function on the left axis, given as 
initial condition the $N$ exact values on the right axis.
All in all, the method provides a reasonable representation of the 
function from the solution of the Cauchy-Riemann equations. Note the 
imaginary part, exactly zero on the left axis, is calculated to be of 
order 1\% with a forty-point grid. This should be considered the error 
of the method, and is far less than the statistical errors in the 
lattice data that we will shortly employ.

Should the analytic structure of the function become available,
one could devise an arbitrary path in the complex plane from the region 
where the function is known to the region where it is wanted by 
analytical continuation. The initial value problem can then be 
formulated with the advance direction along the tangent vector to 
the path, $\mathbf{\tau}$. This vector changes in principle orientation, 
so one would need to use the ``Cartesian-like'' formulation to advance
and the ``polar-like'' formulation to rotate the direction of advance, 
in alternate steps.

%%%%%%%%%%%%%%%%%%%%%%%%%%%%%%%%%%%%%%%%%%%%%%%%%%%%%%%%%%
\subsection{Wick rotation of the quark mass function}
%%%%%%%%%%%%%%%%%%%%%%%%%%%%%%%%%%%%%%%%%%%%%%%%%%%%%%%%

In the introduction we have advanced, in figure 
\ref{fig:minkowski_mass}, our main application,
the analytical continuation of the quark mass function $M(p^2)$ in 
Landau gauge, that we now carefully analyze and justify.

Because of Lorentz invariance applied to a spin $1/2$ fermion, the quark 
propagator can be written in full generality as
\be
S(p) = \frac{iZ(p^2)\not p}{p^2-M^2(p^2)} + 
\frac{iZ(p^2)M(p^2)}{p^2-M^2(p^2)} \ .
\ee
As already discussed, the pole in the denominator is a nuisance usually 
disposed of by performing the Wick rotation $p_0 \to ip^0_E$. 
Concentrating on the denominator then
$$
\frac{1}{p_0^2-{\mathbf{p}}^2-M^2(p_0^2-{\mathbf{p}}^2) } \to
\frac{-1}{p_0^2+{\mathbf{p}}^2+M^2(-p_0^2-{\mathbf{p}}^2) }
$$
the pole is absent for real $M$. One usually eschews a sign, the 
function $M^2(-p_0^2-{\mathbf{p}}^2)$ is customarily called  $M(p^2_E)$ 
where $p^2_E>0$, and we will keep this notation. Hence to retrieve the 
mass function that actually appears in the Minkowski space propagator 
for positive virtuality $p^2$, we need to identify it with the 
analytical continuation of the lattice (or DSE) $M(p^2_E)$ to negative 
$p^2$.
Note that although the analytical continuation is nominally made in 
$p_0$, since $p_0^2-{\mathbf{p}}^2$ is a polynomial, it is an analytical 
function of $p_0$. We conclude that wherever $M$ is analytical in $p_0$ 
it is also analytical in $p_0^2-{\mathbf{p}}^2$ (conversely, there 
needs to be a branch cut in the $p_0$ plane that is absent in terms of 
the Lorentz invariant variable).

The analytical properties of $M(p^2)$ are not well known, especially for 
confined quanta such as quarks. However we note that a pole in this 
function would imply a zero in the quark propagator, and this, assumed 
continuous, can be ruled out in the entire region of the complex plane 
that we sample, from the lattice data on the Euclidean real $p^2$ axis.

We take the lattice data from \cite{Parappilly:2005ei}. 
This has been provided to us from $28^3\times 96$ lattices  with MILC 
configurations. We have used $a^{-1}=2.29\ GeV$ to set the scale from 
the internodal spacing.

\begin{figure}
\includegraphics[width=7cm,angle=-90]{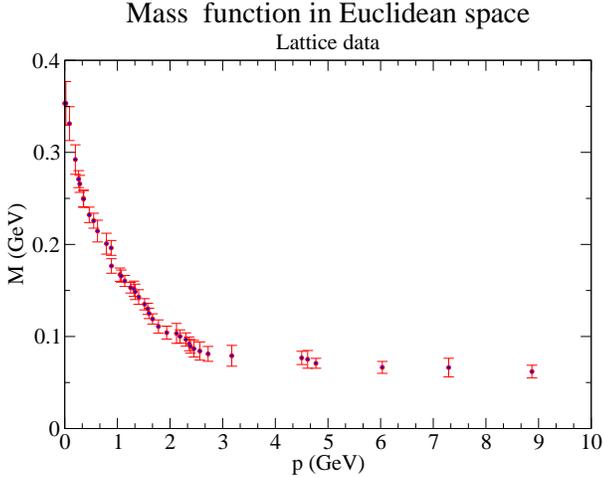}
\caption{\label{fig:lattice}
A part of the lattice data for the quark mass function in Landau-gauge 
QCD. We have trimmed the data to ensure the monotonous decrease in the 
function and reduce the high-frequency noise, which grows fast in 
the Cauchy-Riemann equations. We slightly increased the error bands to 
cover the omitted points (trading our systematic error in trimming 
into a statistical error). The lattice data from 
\cite{Parappilly:2005ei} is normalized to a current quark mass $m_u=60\ 
MeV$.
}
\end{figure}

\begin{figure}
\includegraphics[width=7cm,angle=-90]{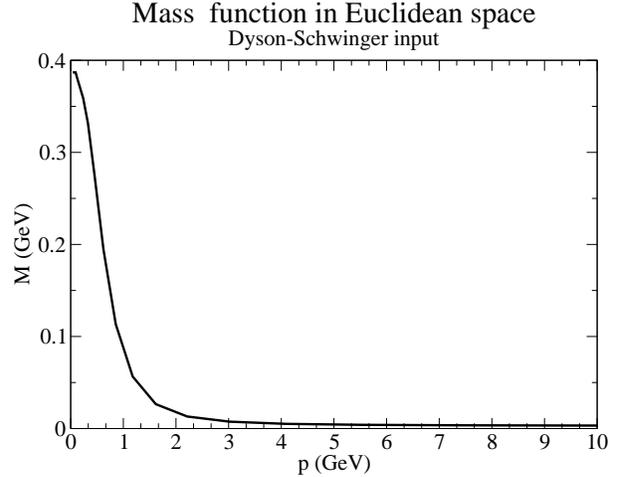}
\caption{\label{fig:DSEinput}
Input calculation from the Dyson-Schwinger formalism. The current quark 
mass is $2\ MeV$ at $13\ GeV$. The function is smoother than lattice 
data, reducing high-frequency noise, in exchange its systematic errors 
are more difficult to control. The original 
function reported in  \cite{Alkofer:2008tt} decreases at very small 
momentum, we have been conservative and avoided this by a small 
variation of the vertex dressing functions.} \end{figure}

Due to our sensitivity to large frequency noise, we only use a subset of 
the lattice data, taking one of every few points in the interval 
$p\in(0,8)\ GeV$. The trimming has been performed so that the resulting 
mass monotonously decrease towards higher momenta (asymptotic freedom), 
to avoid distortions of analyticity and large errors through rapidly 
varying derivative. 
We further square the abscissa $p\to p^2$, since the latter is the 
variable in which we perform the analytical continuation. 
The original lattice data, already in physical $GeV$ units, is given in 
figure \ref{fig:lattice}. We also show the actual input set to our 
code after these manipulations have been performed, that is faithful to 
the original data and error bands, but amenable to analytical 
continuation.
The data is renormalized in the $MOM$ scheme, and as can be seen the 
mass at a scale of $8\ GeV$ is $60\ MeV$ with the chosen scale $a^{-1}$.
We further take as input  a Dyson-Schwinger calculation from 
\cite{Alkofer:2008tt}. This data has as an advantage that there are no 
statistical errors and the function is very smooth. In exchange, there 
are systematic errors (coming from the precise way in which the 
quark-gluon vertex is treated in that reference), that are unknown and 
only controllable in the propagator in comparison with lattice data or 
renormalization group equations. We plot the resulting set in figure
\ref{fig:DSEinput}.

Finally, we perform the analytical continuation on a strip in the 
complex plane above the axis, that presumably avoids 
non-analyticities in $M(p^2)$ (else a continuation under the axis is 
possible) and obtain the real part of $M$ advanced in the introduction 
in figure \ref{fig:minkowski_mass}.

\begin{figure}
\includegraphics[width=7cm,angle=-90]{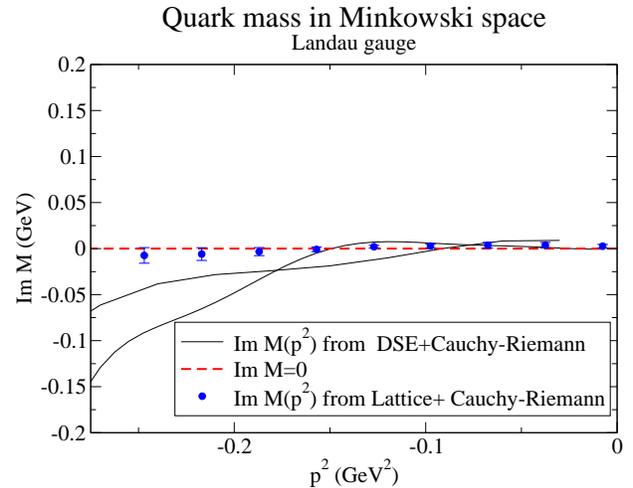}
\caption{\label{fig:mink_massim}
The Cauchy-Riemann method leads to an imaginary part of the 
mass-function on the left half-axis in Euclidean space that is 
well-compatible with $0$. This is in agreement with the outcome of a 
Taylor expansion around the origin (however there is no telling the 
convergence radius of such a series, so we deem the Cauchy-Riemann 
method superior). } 
\end{figure}

We also plot in figure \ref{fig:mink_massim} the imaginary part of the 
same $M$ function, that as can be seen is compatible with zero within 
the error bands. 

From the graphs one can conclude that, just as for the tree-level 
propagator in perturbation theory, there is a crossing of $M$ and $p$ 
for (negative Euclidean), positive Minkowski $p^2$. This means that the 
actual quark propagator does have a pole at or very near the real axis. 
It has been quoted  \cite{Alkofer:2003jj} at $(300-500)\ MeV$ from the 
Dyson-Schwinger equations alone. From the analysis of the lattice data 
set at hand, we conclude $M(M)=305(25)\ MeV$, in agreement with that 
estimate. Of course, it would be interesting to compare with other 
lattice data sets, and in particular use different current quark masses, 
so the error band is definitely larger.

We do not find support for the attending conjecture of two conjugate 
poles with a sizeable imaginary part.

%%%%%%%%%%%%%%%%%%%%%%%%%%%%%%%%%%%%%%%%%%%%%
\section{Some theoretical issues\label{sec:theory}}
%%%%%%%%%%%%%%%%%%%%%%%%%%%%%%%%%%%%%%%%%%%%

%%%%%%%%%%%%%%%%%%%%%%%%%%%%%%%%%%%%%%%%%%%%%%%%%%%%%5
\subsection{Uniqueness}
%%%%%%%%%%%%%%%%%%%%%%%%%%%%%%%%%%%%%%%%%%%%%%%%%%%%%

Here we study to what extent the solution of the Cauchy-Riemann 
equations for the quark mass function is unique, given the initial 
conditions as the lattice computation on the positive $p^2_E$ half-axis.
By standard complex analysis, the uniqueness of an analytic continuation 
of a function is guaranteed if the function is initially known on an 
open subset of $\mathbf{C}$.

The positive real half-axis is open in $\mathbf{R}$, but not in 
$\mathbf{C}$. However it is easy to show that the analytic continuation 
is unique. Imagine that $u$ and $v$ are known for $y=0$ and $x>0$. Then,
all partial derivatives $\frac{\partial^n u}{x^n}$ and $\frac{\partial^n 
v}{x^n}$  are known. In particular for the quark mass function, 
$v(y=0)=0$ (and all $x$-derivatives also vanish), and $u(x,y=0)=M(x)$, 
the real mass. 

Assume that the extension of $M$ to the complex plane was not unique. 
Then in addition to $f=(u,v)$ there would be another function, $f+g$, 
that would satisfy the Cauchy-Riemann equations with the same initial 
conditions. Since the sum of two analytic functions is analytic, 
$g$ itself should be analytic. This means that its components 
$g_x$, $g_y$ would also  satisfy the 
Cauchy-Riemann equations
\ba \label{CRg}
\frac{\partial g_x}{\partial y} = -\frac{g_y}{\partial x} \\
\frac{\partial g_y}{\partial y} =  \frac{g_x}{\partial x} 
\ea
with initial condition $g(y=0)=0$ exactly, with all derivatives
$\frac{\partial^n g(y=0)}{\partial x^n}=0$ also vanishing on the real 
axis. Automatically, employing eq. (\ref{CRg}), and subsequently 
deriving it, all $y$ derivatives also vanish. Therefore $g$ is exactly 
zero in the domain of analyticity, and $f$ unique.

Of course, in practice $f$ is only known at a discrete and finite set of 
points $z_i=(x_y,0)$ $i=1..N$. An analytic function could oscillate 
between any two of the points and take arbitrarily large or small 
values.  Therefore one needs an additional hypothesis to claim that the 
computed function is a fair representation of the ``actual'' function. 

The sufficient hypothesis is monotony of the function between any two 
grid points (note the function might be globally non-monotonous by 
being allowed to change derivative sign at the grid points themselves). 
If the function is strictly 
decreasing between $x_i$ and $x_{i+1}$, then the maximum and minimum 
values that it can take between them are $f_i$ and $f_{i+1
}$, and the 
function is bound (it being analytic, it is also continuous). 
Then, to arbitrarily shrink the error in our knowledge of the initial 
condition, one just needs to arbitrarily shrink the grid spacing, so as 
to further constrain the function in every subinterval. The function 
computed with the discretized Cauchy-Riemann equations will be as close
to the true function as the stability of the system allows, given the 
bound error in the initial conditions.

For our example, the light quark propagator, there is essentially no 
question that the mass function is monotonously decreasing towards 
larger momenta. This is known at large momentum from asymptotic QCD and 
at low momentum from all studies of Dyson-Schwinger equations and 
lattice (where all non-monotonous behavior has way less than $1\sigma$ 
significance and can be safely called noise). The hypothesis of monotony 
can be checked (falsified) with 
lattice data by simply decreasing the link size in the grid while at 
the same time reducing the statistical error bar.
 
  One more caveat can be raised.  Imagine adding to the ``actual'' 
function $f(u,v)$ another analytical function $g(u,v)$ such that 
$g$ is very near zero on the right (Euclidean) axis and very large on 
the left hand (Minkowski) side. Then, while $f+g$ does not exceed the 
error bars for $f$ on the initial data, it completely changes the answer 
on output since $f+g$ is very different from $f$ on the right half-axis.
  Of course, the derivative of the function must be very large
 around $u=0$ since the function changes from very small to 
sizeable in a small interval. To bind this derivative from above and 
exclude this unpleasant possibility one needs to demand an additional 
condition, since exact knowledge of all derivatives of the function or 
knowledge of the function in the entire interval is, 
in a computer grid representation, unavailable. 

Now, a fast change of the derivative beyond what is visible from the 
data points implies that the second derivative is not well represented 
by its discrete approximation. 
Here one can demand monotony of the second derivative of the function 
between the last three (few) points of the grid on the interval $x_1$, 
$x_2$, $x_3$. This guarantees that the extrapolation of the derivative 
at just  the last grid point does not grow arbitrarily, since the 
second derivative remains bound. This is now 
quite a technical condition, and maybe not optimal, others being 
possible.

%%%%%%%%%%%%%%%%%%%%%%%%%%%%%%%%%%%%%%%%%%%%%%%%%%
\subsection{Instability of large frequency noise \label{subsec:stab}}
%%%%%%%%%%%%%%%%%%%%%%%%%%%%%%%%%%%%%%%%%%%%%%%%%%

Let us now consider the effect of a perturbation on the system of
Cauchy-Riemann equations (\ref{CRg}).
Since the system is linear, it accepts a Fourier analysis. Let us 
perform it on the $x$ variable so that the Fourier components are
\ba
g_x(x,y)= A(k,y) e^{ikx}\\ \nonumber
g_y(x,y)=B(k,y)  e^{ikx} \ .
\ea
Then the system of equations becomes
\be
\frac{\partial}{\partial y} \left(
\begin{tabular}{c}
$A$ \\ $B$
\end{tabular} \right) =
\left[ \begin{tabular}{cc}
$0$ & $-ik$ \\
$ik$ & $0$ \end{tabular}\right]
\left( \begin{tabular}{c}
$A$ \\ $B$
\end{tabular}
\right)
\ee
and therefore
\be
\left( \begin{tabular}{c}
$A$ \\ $B$
\end{tabular} \right) =
\left[ \begin{tabular}{cc}
$\cosh ky$ & $-i\sinh ky$ \\
$i\sinh ky$ & $\cosh ky$ \end{tabular}\right]
\left( \begin{tabular}{c}
$A_0$ \\ $B_0$
\end{tabular}
\right) 
\ee
in terms of the initial condition on the $x$-axis.
Obviously, if the exact solution is initially perturbed due to computer 
inaccuracies by an amount $\delta A_0$, with $B_0=\delta B_0=0$ for 
simplicity, then at large distances the perturbation on the computed 
solution exponentiates
$$
\delta A\ ,\ \ \delta B \propto e^{ky} \ar \delta A_0 \ar \ .
$$

This could of course be anticipated by remembering that the solutions of 
the Cauchy-Riemann equations are two-dimensional harmonic functions, so 
that the separable solutions are sinusoidal functions times 
exponentials, $\cos kx \cosh ky$, etc.

\begin{figure}[htbp]
\includegraphics[width=7cm,angle=-90]{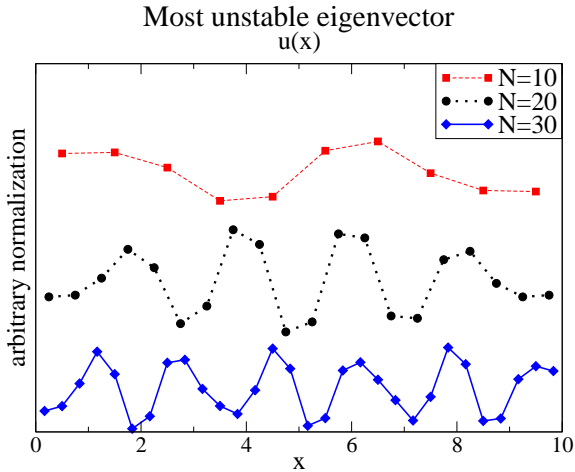}
\caption{The unstable most eigenvector as a function of the number of 
points transverse to the direction of advance, for the iteration matrix 
in the Cauchy-Riemann  $\theta=-0.5$ discretization. 
The functions have been vertically shifted for visibility.
 \label{fig:unstable}}
\end{figure}

This is especially worrysome when using Montecarlo data for the initial 
condition, since the local (high-$k$) noise spoils stability very soon. 
Thus, one needs to apply a cooling algorithm or trim the data first to 
remove short-distance fluctuations, justifying our keeping only part of 
the lattice data to ensure monotony. Large-distance, systematic shifts 
of the initial condition are less perturbing.
In figure \ref{fig:unstable} we capture the $u$ (real part) 
noisiest eigenvector of the iteration matrix 
${\mathbf{A}}^{-1}\mathbf{B}$
for a particular $\theta=-0.5$ method for fixed grid size, to show its
increasing wavenumber.

The iteration matrix ${\mathbf{A}}^{-1}\mathbf{B}$ always has 
eigenvectors that are larger than 1 in modulus. We have studied them for 
very simple equispaced rectangular grids, with the direction of advance 
along $y$. For fixed ratio of the increment $\frac{\Delta y}{\Delta x}$, 
the largest eigenvalue is quite independent of the number of grid points 
(obvious from the definition of the matrix). The dependence with 
$\theta$ and $\frac{\Delta y}{\Delta x}$ can be followed from table 
\ref{tabla:autovalores}
\begin{center}
\begin{table}\caption{ Largest eigenvalue (one of a pair) for the 
iteration matrix of the $\theta$ method. 
\label{tabla:autovalores}
}
\begin{tabular}{cc|c}
$\theta$ & $\frac{\Delta y}{\Delta x}$ & $\ar \lambda \ar $ \\ \hline
2.5   & 1 & 9.8 \\
2.0   & 1 & 2.6 \\
1.5   & 1 & 11 \\
1.0   & 1 & 13 \\
0.5   & 1 & 2.7 \\
0.1   & 1 & 2.0 \\
-0.1  & 1 & 1.8 \\
-0.5  & 1 & 1.6 \\
-1.0  & 1 & 11 \\
-1.5  & 1 & 13 \\
-2.0  & 1 & 2.7 \\
-2.5  & 1 & 7.8 \\
\hline
-0.5 & $\frac{1}{2}$  & 1.4 \\
-0.5 & $\frac{1}{4}$  & 1.2 \\
-0.5 & $\frac{1}{6}$  & 1.14 \\
-0.5 & $\frac{1}{8}$  & 1.11 \\
-0.5 & $\frac{1}{10}$ & 1.08
\end{tabular}
\end{table}
\end{center}
Note from the table that, unlike for the heat equation, the $\theta$ 
method is not convergent. We are not able to approach a given function 
with arbitrary accuracy, but only provide an estimate. It is apparent
that decreasing the advance step $\frac{\Delta y}{\Delta x}$ is not a 
winning strategy, since, for example, fixing $\theta=-0.5$, with
$\frac{\Delta y}{\Delta x}=1$, $\ar \lambda \ar = 1.6$ and to advance 
the same distance with $\frac{\Delta y}{\Delta x}=1/10$, one needs ten 
steps, but $1.08^{10}=2.1>1.6$, meaning that with a smaller step, one 
can advance less far in the progress direction since errors amplify 
faster. (Of course, by decreasing the step one does obtain a more 
reliable representation of the function for short advance distances).

%%%%%%%%%%%%%%%%%%%%%%%%%%%%%%%%%%%%%%%%%%%%%%%%%%%%%%%%%%%%%%%%%%%%
\subsection{Generalization to several complex variables}
%%%%%%%%%%%%%%%%%%%%%%%%%%%%%%%%%%%%%%%%%%%%%%%%%%%%%%%%%%%%%%%%%%%
In principle it would appear straightforward to generalize the Wick 
rotation to several dimensions. For example, let us consider a vertex 
function in field theory, say the quark and gluon or the 
electron-photon three-point functions. These are characterized by twelve 
Dirac tensors multiplied by amplitudes of the three independent Lorentz 
scalar variables, the squared momenta of each of the particles,
$$
\lambda_i(p_1^2,p_2^2,q^2)\ \ \ i=1..12  \ .
$$
Given the lattice data in Euclidean space, 
$$
\lambda_i(p_{1E}^2,p_{2E}^2,q_E^2)\ \ \ i=1..12  
$$
one would need to perform the inverse Wick rotation to negative 
$p_{E}^2$ in each of the variables, in practice solving the 
Cauchy-Riemann equations variable by variable. Notice if the power-law 
solutions of \cite{Alkofer:2008tt} are correct, then one expects a cut 
at zero virtuality $q^2=0$ in the gluon variable, but this can be avoided 
by appropriately deforming the region where one solves the 
Cauchy-Riemann equation.

Now let us examine a curiosity that does not come about in one 
dimension. If there is only one variable, $p^2=0$ defines a light-cone 
in Minkowski space, a three-dimensional manifold in four-dimensional 
space. However, $p_E^2=0$ defines the origin in Euclidean space, just a 
point (this is just another manifestation of the difference between the 
compact rotation group and the unbound Lorentz group).
The interesting observaion is that, upon Wick-rotation, $f(p_E^2=0)$, 
the value of a Green's function at one point in Euclidean space, becomes
$f(p^2=0)$, the value of the same Green's function on the entire 
light-cone.

But what happens in more dimensions? One may know the value of the 
function $\lambda(p_{E1}^2=0,p_{E2}^2,p_{E2}^2)$ in Euclidean space 
at the origin for the variable $p_{E1}$.
But in Minkowski space $p_1^2=0$ does not imply $p_1=0$, hence
the function takes different values for different points of the $p_1$ 
light-cone, $\lambda(p_1^2=0,p_2^2,(p_1-p_2)^2$ 
that do not coincide with the value at the origin in Euclidean space.

This comes about because in a three-point function there are two 
reference four-vectors, $p_1$ and $p_2$, and while the Euclidean point 
with $p_1=0$ is at a fixed distance from the poin

%%%%%%%%%%%%%%%%%%%%%%%%%%%%%%%%%%%%%%%%%%%%
\subsection{Taylor expansion}
%%%%%%%%%%%%%%%%%%%%%%%%%%%%%%%%%%%%%%%%%%%

Analytical functions accept Taylor expansions of the type
\be \label{Taylor}
f(z)= \sum_{n=0}^\infty \frac{(z-z_0)^n}{n!} f^{(n)}(z_0)\ .
\ee
One could think of performing a polynomial fit to a given set of data 
points $(z_i,f(z_i))$ to represent the function within the radius of 
convergence of the series. One would simply need to solve the system of 
Vandermonde for $N$ points
\be \label{Vandermonde}
f(z_i)=\sum_{n=0}^N (z-z_0)^N \left( \frac{f^{(n)}(z_0)}{n!}\right) \ .
\ee
For example, expanding around the origin, the matrix of coefficients for 
the linear system is the Vandermonde matrix with rows
$(1,z_i,z_I^2,\dots z_i^N)$. Once the system has been solved for  the 
derivatives $\left( \frac{f^{(n)}(z_0)}{n!}\right)$, they can be 
substituted in eq. (\ref{Taylor}) to obtain the function at an arbitrary 
point.

One would argue that this is the simplest local implementation of 
analyticity, and why should one worry about the Cauchy-Riemann equations 
at all. Of course, in practice monotony is difficult to achieve with a 
finite number of polynomials: the approximant will oscillate between the 
tabulated grid points with the lattice data. In addition, should there 
be a cut starting at $p^2=0$ in the complex plane, that would not be 
surprising in view of the power-law representations reported in the 
literature for other Green's functions  \cite{Alkofer:2003jj},
the radius of convergence of the Taylor series would be exactly zero.  
Although such difficulties can be circumvented, a
practical implementation would become as or more difficult than the 
Cauchy-Riemann equations. We have not pursued the matter further.

%%%%%%%%%%%%%%%%%%%%%%%%%%%%%%%%%%%%%%%%%%%%%
\section{Summary and conclusions  \label{sec:conclusions}}
%%%%%%%%%%%%%%%%%%%%%%%%%%%%%%%%%%%%%%%%%%%%

We have presented a first analysis of the Cauchy-Riemann equations as 
applied to performing the inverse Wick rotation from Euclidean to 
Minkowski momenta. Given that their square is the elliptic Laplace 
equation, and that they are not of variational type \cite{ferreiro}, we 
do not have a trivial method to solve them to arbitrary accuracy. 
However they 
provide an estimate of a function by analytical continuation as an 
initial value problem, that relies only on the existence of a path 
on the complex plane that is free of singularities.
While much mathematical literature exists \cite{reffullerton} and more 
sophisticated solution methods have been reported, the very elementary 
methods presented here are sufficient to gain insight into the quark 
propagator mass function. 

We obtain, after analytical continuation of lattice QCD data, 
qualitatively  backed up by a Dyson-Schwinger calculation, and
 in agreement with recent studies, a pole of the quark propagator in 
Landau gauge QCD at or 
very near the real axis, with a mass of $305(25)\ MeV$. We find no 
support for the sometimes conjectured pair of complex conjugate poles.
\vspace{1cm}

%%%%%%%%%%%%%%%%%%%%%%%%%%%%%%%%%%%%%%%%%%%%%%%%%%%%%%%
\emph{The authors thank P. Bowman for his kind explanations and
providing them with the quoted lattice data. Felipe Llanes-Estrada
thanks the team at the Instituto Superior Tecnico de Lisboa for 
their hospitality during parts of this project.
Mercedes Gimeno-Segovia thanks the ``Excellence Scholarship
Programme'' of the Comunidad de Madrid for financial support.
Work supported  by grants FPA 2004-02602, 2005-02327,
PR27/05-13955-BSCH \\
(Spain) and Acci\'on Integrada Hispano-Lusa
HP2006-0018.}
%%%%%%%%%%%%%%%%%%%%%%%%%%%%%%%%%%%%%%%%%%%%%%%%%%%%%%

%%%%%%%%%%%%%%%%%%%%%%%%%%%%%%%%%%%%%%%%%%%%
\appendix
\section{\\ Lorentz invariant discretizations in Minkowski space}
%%%%%%%%%%%%%%%%%%%%%%%%%%%%%%%%%%%%%%%%%%%%
Physical states provide representations of the rotation group. In 
lattice gauge theory, the symmetry of the grid reclassifies the states 
but at least offers some control over what signal may belong to what 
spin multiplet. This is because the lattice is invariant under a 
subgroup of the rotation group, typically tagged by a minimum angle 
$\theta=90$ degrees. Likewise, the grid is invariant under a subset of 
the translation group characterized by the grid spacing $a$.

However there is no equivalent parameter for finite grids in Minkowski 
space. A grid that is invariant under discrete Lorentz transformations 
has infinitely many points (and is therefore not tractable on a 
computer). We discuss 
this shortly as one more advantage of working in Euclidean space that 
can come to play when computing pdf's or reducing Green's functions in 
non equal-time gauges such as the light-front gauge.

To see it, it is easiest to observe that Lorentz 
transformations map  the 
light-cone to itself, acting as dilatations within this manifold. To 
see it, it is simpler to work in 1+1 dimension, where the light-cone 
becomes the pair of lines
$$
x=t;  \ \ x=-t \  .
$$
The action of the Lorentz boost becomes
\begin{equation} \label{tLorentz}
\left( \begin{tabular}{c} $t_1$ \\ $x_1$ \end{tabular} \right)
=
\left( \begin{tabular}{cc}
$\gamma$ & $\beta \gamma$ \\
$\beta \gamma$ & $\gamma$
\end{tabular} \right)
\left( \begin{tabular}{l} $t_0$ \\ $x_0=t_0$ \end{tabular} \right)\ ,
\end{equation}
hence
\begin{equation}
x_1=t_1= t_0\ (\gamma +\beta \gamma)\ ,
\end{equation}
a dilatation of parameter $\gamma(1+\beta)$.
There is now an obvious way to construct a discretization of the 
light-cone that is invariant under Lorentz transformations of a fixed 
parameter $\gamma$. Simply pick an arbitrary point $(t_0,x_0)$ and 
obtain the sequence $(t_1,x_1)$, $(t_2,x_2)$... obtained by successively 
applying eq. (\ref{tLorentz}) to it. The infinite sequence so obtained 
is such that every point $j$ is the image of another point $j-1$ under a 
Lorentz transformation, except for $j=0$. To generate this one we 
further need to also include in the discretization all points obtained 
by successively applying to $(t_0,x_0)$ the inverse Lorentz 
transformation
\begin{equation}
\left( \begin{tabular}{c} $t_{-j}$ \\ $x_{-j}$ \end{tabular} \right)
=
\left( \begin{tabular}{cc}
$\gamma$ & $-\beta \gamma$ \\
$-\beta \gamma$ & $\gamma$
\end{tabular} \right)
\left( \begin{tabular}{l} $t_{-j+1}$ \\ $x_{-j+1}=t_{-j+1}$ 
\end{tabular} \right)\ ,
\end{equation}

\begin{figure}[htbp]
\includegraphics[width=7cm]{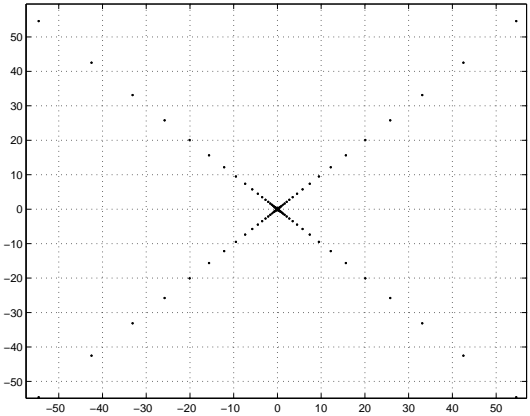}
\caption{A Lorentz transformation acts on the light-cone simply as  
a dilatation. Shown is a discretization invariant under a discrete 
Lorentz transformation, where each point on the plot is the 
Lorentz-transformed of its nearest neighboor towards the origin of 
coordinates. Note this discrete transformation of parameter 
$\Lambda\equiv \gamma(1+\beta)$ belongs to a subgroup of transformations 
with parameters $\Lambda^2$...$\Lambda^n$, etc.
 \label{fig:ejes}}
\end{figure}

Thus, it appears that we have a simple discretization of the line $x=t$ 
that is invariant under discrete Lorentz transformations (dilatations in 
this line). It is however not invariant under translations, as the 
spacing between points $x_j-x_{j-1}\ \ j>0$ increases in proportion to 
their distance to the origin. Therefore, as is well known, one cannot 
construct a web that is simultaneously translation and Lorentz 
invariant, even under discrete Lorentz transformations.

This is different from the Euclidean space case, where the 
translationally-invariant Bravais lattices are also invariant under the 
discrete rotations of their corresponding crystallographic point group.

Moreover, grids that are invariant under a discrete Lorentz 
transformation, have infinitely many points dense at the origin and 
infinity.

To construct a discretely Lorentz-invariant lattice of the entire 1+1 
dimensional space, we just have to write-down a discretization 
in which every point is the Lorentz transformed of another (and every 
Lorentz-transformed point belongs to the lattice). 
Since Lorentz transformations leave the metric invariant, 
the hyperbolae $k^2=(k^0)^2-({\bf k}^2)=m^2$ are invariant, that is, it 
is sufficient to construct a discretization of the hyperbola of  (mass)
parameter $m$. For this, all one needs to do is to choose a point 
$y_0$ on the hyperbola and the discrete Lorentz transformation of 
parameter $\gamma$, and apply the Lorentz transformation and its 
inverse to generate the sequence of points 
$$
y_n= \Lambda(\gamma)^n y_0 \ \ \ y_{-n}= \Lambda(\gamma)^{-n} y_0 \ .
$$
Once this has been achieved, it is sufficient to pick up a set of 
hyperbolae to cover the entire space at wish and this completes the 
discretization. 

To construct it, we start with a $y_0$-lattice in wich there are points 
over both coordinate ($x$, $t$)axes. Then we obtain two families of 
hyperbolae with equations
\ba \label{no light cone}
H_1=\frac{x^2-(ct)^2} {a_i^2}  =1 \\ \nonumber
H_2=\frac{(ct)^2-x^2}{a_i^2} =1
\ea
where $a_i=ia$ is the distance between their vertices (which will 
coincide with the chosen $y_0$) and the coordinate origin.

Note that the fact that a point and its Lorentz-transformed belong 
to $H_1$ or $H_2$ depends on whether the point lies within the 
forward or backward light-cones (hyperbolae of type $H_2$) or not 
(hyperbolae of type $H_1$)
Note also that the successive Lorentz-transformed images of a point 
under a discrete boost do not fill the hyperbola, they define a 
discrete lattice over it. All that remains is to find  this subset of 
points.

A geometric way to map the discretization of one
hyperbola to all others is to simply take straight lines through every
point on the lattice covering the reference hyperbola and the origin.

This family of straight lines follow either of the rules (depending on 
the sign of the discrete Lorentz transformation employed in the 
construction):
\begin{eqnarray} \label{firstfl}
(ct)=(\tanh(i\varphi))x  \\  \nonumber
(ct)=\frac{1}{\tanh(i\varphi)}x
\end{eqnarray}
where $\varphi$ is the discrete hyperbolic angle labeling the boost, and 
$i$  the number of direct (positive $i$) or inverse (negative $i$) 
Lorentz transformations of the initial vertex.

\begin{figure}[h]
\includegraphics[width=7cm]{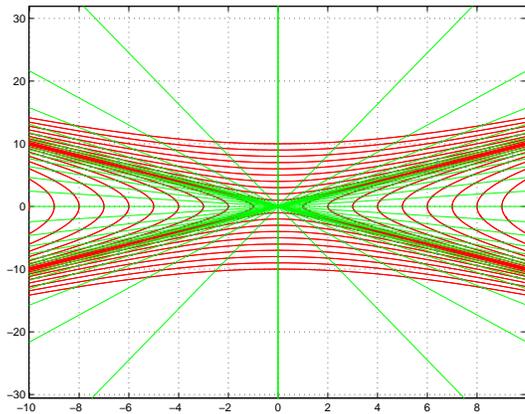}
\caption{	\label{fig:redcompleta}
The intersection points of a family of hyperbolae and a family of 
straight lines define a Lorentz-invariant lattice of 1+1 dimensional 
Minkowski space. The number of points in the lattice is infinite for any 
discrete set of boosts of parameter $\varphi$. 
 }
\end{figure}

Combining eq. (\ref{no light cone}), and eq. 
(\ref{firstfl}), we finally construct a Lorentz 
invariant lattice in (1+1) dimmension of a chosen $\varphi$ parameter 
at the intersections of the two families. The situation is plotted in 
figure \ref{fig:redcompleta}.

Now, it's easy to observe that the hyperbolae contain all the images 
under direct and inverse Lorentz transformations of the vertex $y_0$, 
while the family of straight lines connect the points that correspond to
the $\Lambda^i$ image  of each  vertex across a family of hyperbolae.

%%%%%%%%%%%%%%%%%%%%%%%%%%%%%%%%%%%%%%%%%%%%%

%%%%%%%%%%%%%%%%%%%%%%%%%%%%%%%%%%%%%%%%%%%%

\end{document}